\documentclass[11pt]{article}
\usepackage{url,color}

\usepackage{graphicx}
\usepackage{natbib}
\usepackage{amsmath}
\usepackage{amssymb}
\usepackage{multirow}
\usepackage{fullpage}
\usepackage{setspace}
\usepackage{times}
\usepackage[colorlinks = true,
            linkcolor = black,
            urlcolor  = blue,
            citecolor = black,
            anchorcolor = black]{hyperref}
\date{}
\usepackage[table]{xcolor}% http://ctan.org/pkg/xcolor

\newcommand{\argmax}{\operatorname*{arg \ max}}

\newcommand{\bbeta}{\boldsymbol{\beta}}
\newcommand{\bsigma}{\boldsymbol{\sigma}}

\begin{document}

% Title of paper
\title{Gaussian process regression for survival time prediction with genome-wide gene expression}

% List of authors, with corresponding author marked by asterisk
\author{Aaron J. Molstad, Li Hsu, and Wei Sun\footnote{Correspondence: wsun@fredhutch.org}\\
 Biostatistics Program, Fred Hutchinson Cancer Research Center}
% Author addresses

% E-mail address for correspondence

% % Running headers of paper:
% \markboth%
% % First field is the short list of authors
% {A.J. Molstad, L. Hsu, and W. Sun}
% % Second field is the short title of the paper
% {Genome-wide Gaussian process regression }

% \history{Received August 1, 2010;
% revised October 1, 2010; 
% accepted for publication November 1, 2010}

\maketitle

% Add a footnote for the corresponding author if one has been
% identified in the author list
% \footnotetext{To whom correspondence should be addressed.}

\begin{abstract}
{Predicting the survival time of a cancer patient based on his/her genome-wide gene expression remains a challenging problem. For certain types of cancer, the effects of gene expression on survival are both weak and abundant, so identifying nonzero effects with reasonable accuracy is difficult. As an alternative to methods that use variable selection, we propose a Gaussian process accelerated failure time model to predict survival time using genome-wide or pathway-wide gene expression data. Using a Monte Carlo EM algorithm, we jointly impute censored log-survival time and estimate model parameters. We demonstrate the performance of our method  and its advantage over existing methods in both simulations and real data analysis. The real data that we analyze were collected from 513 patients with kidney renal clear cell carcinoma and include survival time, demographic/clinical variables, and expression of more than 20,000 genes.  
%Our real data analysis incorporates both genome-wide effects and pathway-specific effects. 
Our method is widely applicable as it can accommodate right, left, and interval censored outcomes; and provides a natural way to combine multiple types of high-dimensional -omics data. An R package implementing our method is available for download at \href{http://github.com/ajmolstad/SurvGPR}{github.com/ajmolstad/SurvGPR}. \\
\smallskip

}
\end{abstract}

\onehalfspacing
\section{Introduction}

Predicting the survival time of a cancer patient based on his/her genome-wide gene expression is a well studied, yet unresolved problem. In some types of cancer, the effects of gene expression are both weak and abundant which, when combined with often high censoring rates, makes feature selection for survival time association very challenging. On the other hand, genome-wide gene expression data can be highly informative for prognosis. For example, \cite{zhu2017integrating} demonstrate that two patients with similar genome-wide gene expression data may have similar survival time.

\begin{figure}[t!]
\centerline{\hfill\includegraphics[width=.40\textwidth]{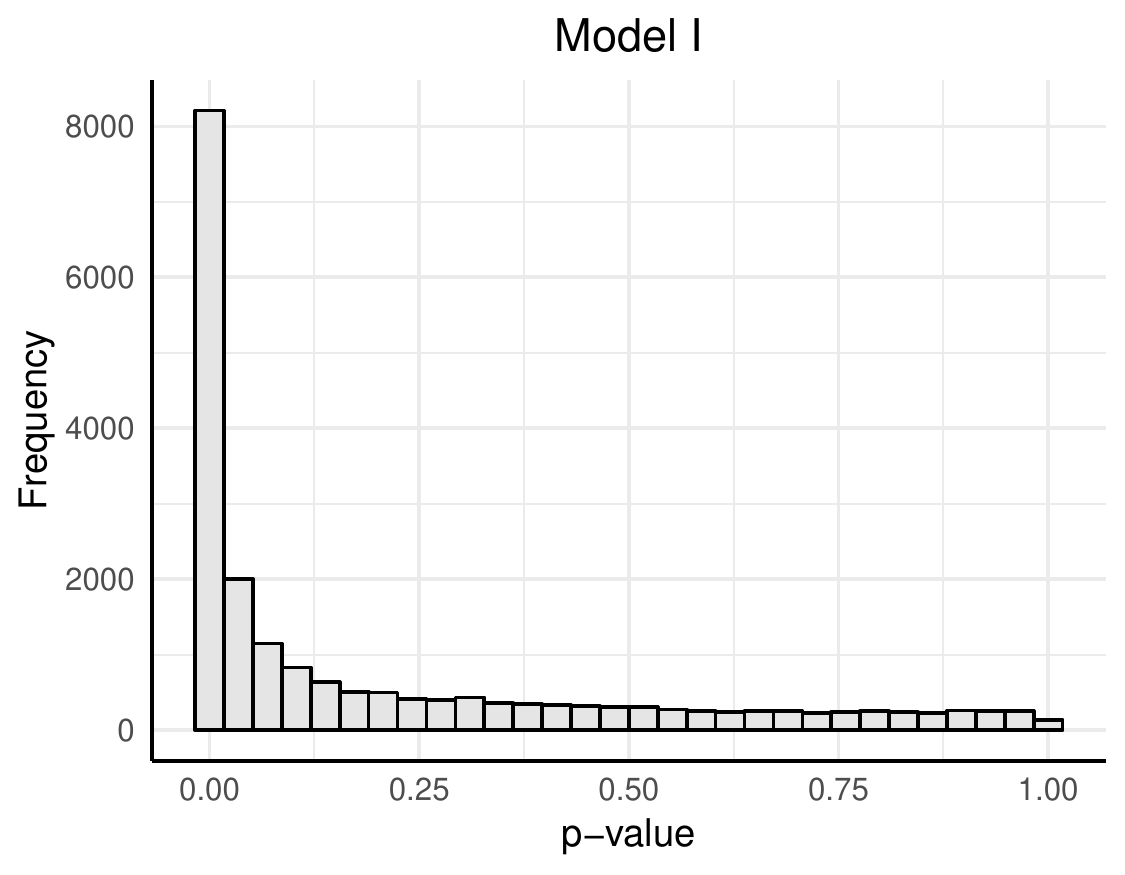}\hfill \includegraphics[width=.40\textwidth]{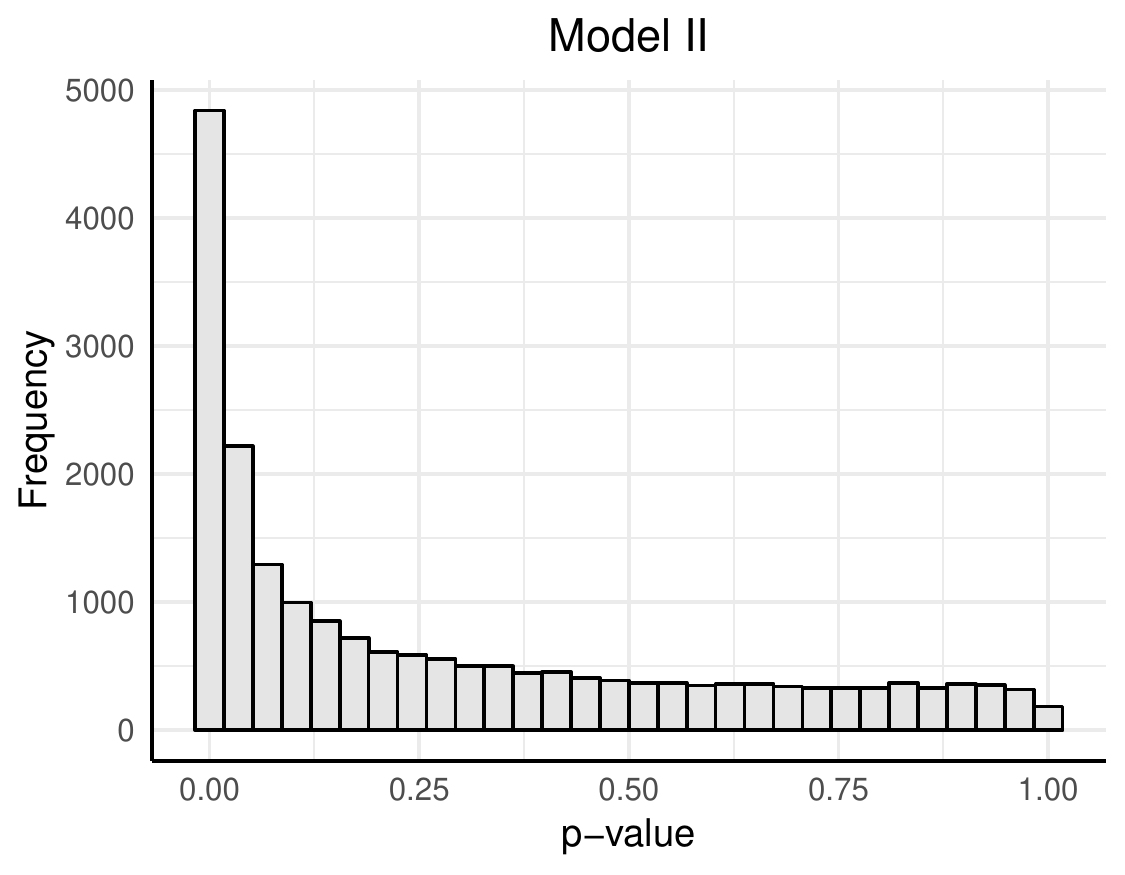}\hfill}
\centerline{\hfill\makebox[.40\textwidth]{(a)}\hfill\makebox[.40\textwidth]{(b)} \hfill}
\centerline{\hfill\includegraphics[width=.40\textwidth]{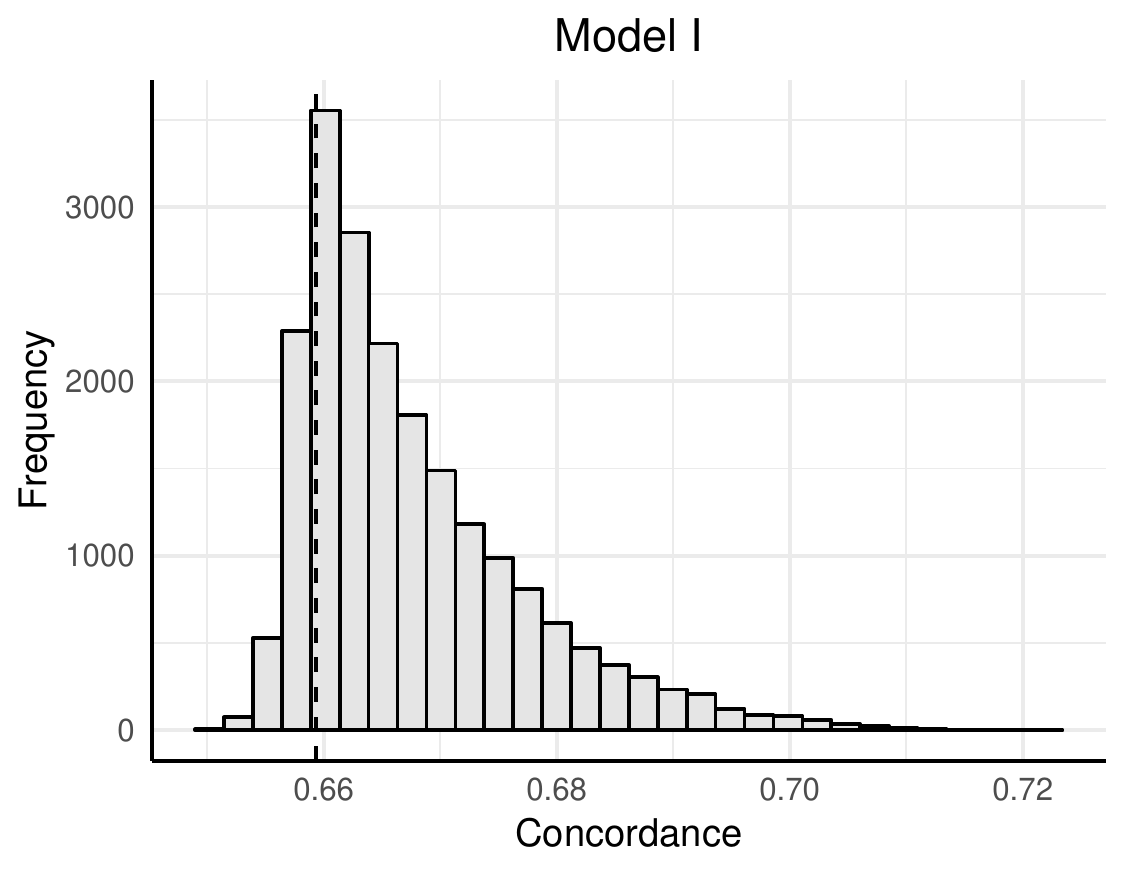}\hfill \includegraphics[width=.40\textwidth]{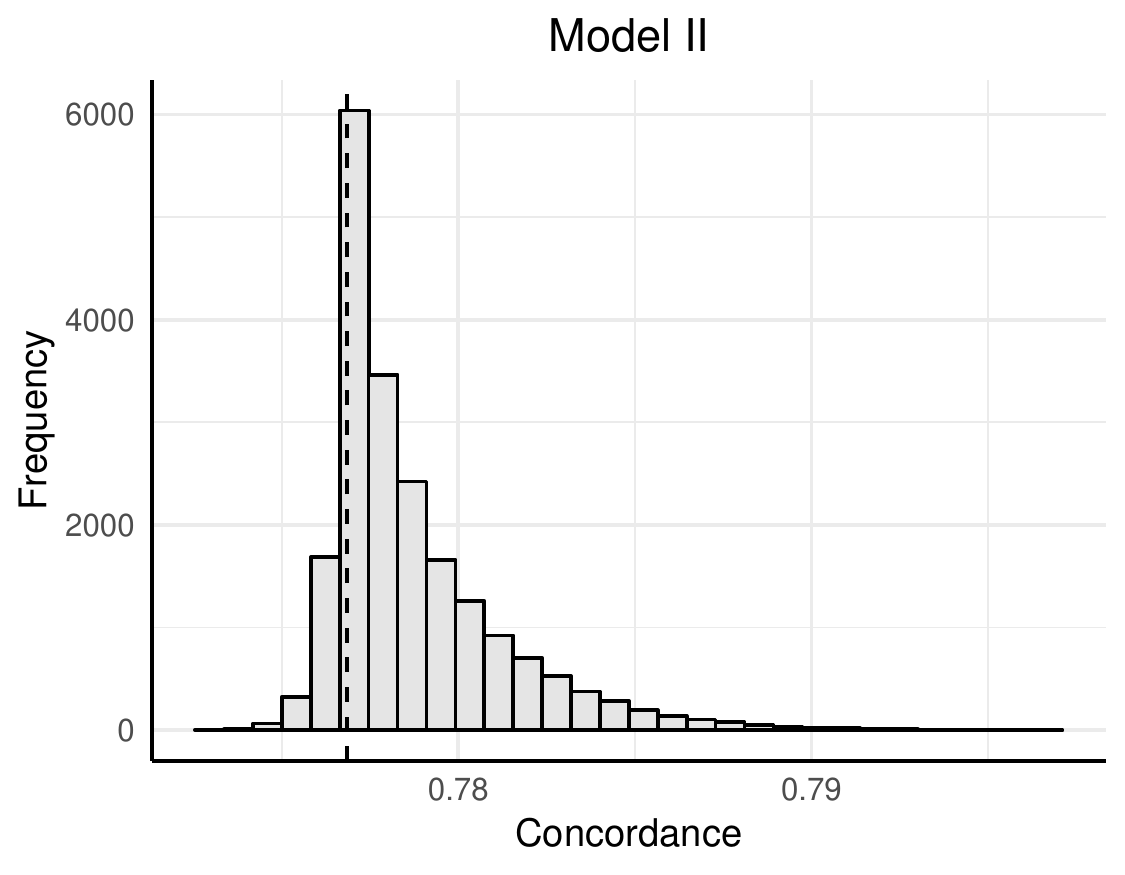}\hfill}
\centerline{\hfill\makebox[.40\textwidth]{(c)}\hfill\makebox[.40\textwidth]{(d)} \hfill}
\caption{Histograms of the marginal p-values for each of the 20,483 genes for (a): Model I and (b): Model II.  Histograms of the concordance for each of the 20,483 marginal models for (c): Model I and (d): Model II. Dark dotted lines in (c) and (d) denote the concordance measurement of survival time prediction under the baseline model that includes all the covariants in Model I or II, other than gene expression. }\label{fig:p_val_dist}
\end{figure}

Our method development is motivated by a dataset with genome-wide gene expression, survival time, and some demographical/clinical variables of more than 500 patients with kidney renal clear cell carcinoma, which is part of The Cancer Genome Atlas (TCGA) project (\href{http://cancergenome.nih.gov/}{http://cancergenome.nih.gov/}). 
% A recent pan-cancer study demonstrates that in kidney cancer, gene expression data are highly informative for prognosis \citep{zhu2017integrating}. In addition, long survival time in kidney cancer has been associated with immune-mediated anti-tumor response \citep{escudier2012emerging}. Thus, studying the effects of gene expression on the survival time in kidney cancer patients may lead to insight on the interactions between tumor cells and the immune system. This is a timely question because kidney cancer patients have unexpectedly high response rates for immunotherapy, given their relatively low mutation burden \citep{yarchoan2017tumor}, and understanding the tumor-immune interaction is crucial for studying the mechanism of immunotherapy response. 
To demonstrate that the associations between gene expression and survival time are abundant and weak in this dataset, we first report results of gene-by-gene marginal association testing. For each gene, we fit two Cox proportional hazards models. In Model I, we include only the expression of this gene as a predictor and the sequencing plate ID as a confounder. In Model II, we include the expression of this gene, sequencing plate ID, and three demographical/clinical covariates: age, gender, and tumor stage. Histograms of the marginal p-values are displayed in Figure \ref{fig:p_val_dist}(a) and (b).

Assuming genes with p-value larger than 0.5 are not associated with survival time, we can calculate the expected number of genes associated with survival time by $p(1 - 2 s/p)$, where $s$ is the number of genes with p-value $>$ 0.5 and $p=20,483$ is the total number of genes. This number is 13,052 and 10,512 for Models I and II, respectively.  Similarly, with a false discovery rate of 0.05, the number of genes that are significantly associated with survival time are 8,550 and 4,312 for Models I and II, respectively. The large number of genes associated with survival time is biologically plausible given that kidney renal clear cell carcinoma is characterized by oncogenic metabolism and epigenetic reprogramming, both of which may affect the expression of many genes  \citep{cancer2013comprehensive}. 

In Figures \ref{fig:p_val_dist}(c) and (d), we display the survival time prediction concordances (C-index) for the 20,483 marginal models. The dark dashed vertical lines denote the concordance of the baseline model, i.e., the model excluding gene expression but including sequencing plate ID (Model I) as well as clinical covariates (Model II). For Model I, including a single gene can improve concordance by as much as six percent. Comparatively, the improvement in concordance for Model II is smaller, with a single gene improving concordance no more than two percent. The concordance improvements indicate that gene expression can improve the prediction of survival time, although few, if any, genes appear to have strong effects. Together, these results suggest that screening or variable selection may be difficult or ineffective because of potentially weak and abundant effects.

%          Prop < .05 FDR < .05 FDR < .10 FDR < .20
%Model I         0.50      8550     10136     12137
%Model II        0.35      4312      6060      8442

% %In Model III, after adjusting for both clinical covariates and surrogate variables, few genes (26) are found to be significantly associated with survival time even when the false discovery rate is set to 0.20. Examining the QQ-plots in Figure 1, it would appear that a large number of genes have weak associations with survival, even after including clinical variables and surrogate variables. 
% The fact that we have little power to identify those $\sim$\textcolor{red}{xxx} genes associated with survival time in model III  demonstrate that screening or variable selection may be difficult or ineffective because of the weak and abundant effects. 

% \begin{table}
% \centering
% \begin{tabular}{r|r|rrr}
%   \hline
%  & Prop $<$ .05 & FDR $<$ .05 & FDR $<$ .10 & FDR $<$ .20 \\ 
%   \hline
% Model I & 0.56 & 10185 & 11579 & 13263 \\ 
%   Model II & 0.35 & 4312 & 6060 & 8442 \\ 
%   Model III & 0.10 & 2 & 5 & 26 \\ 
%    \hline
% \end{tabular}
% \caption{Summary for the p-values for the marginal association screenings between survival times and gene expression in the TCGA KIRC dataset. }
% \end{table}

In our proposed method, we do not attempt to identify a subset of genes associated with survival time. Instead, we use genome-wide gene expression to model the covariance of the log-survival time under a Gaussian process accelerated failure time model. Inspired by multiple kernel learning \citep{gonen2011multiple}, we allow the covariance to be a linear combination of $M$ user-specified candidate kernels. A major challenge for survival time prediction is censoring. To mitigate this challenge, we develop an efficient Monte Carlo EM algorithm which jointly imputes censored log-survival times and estimates model parameters. The imputed survival times are then used in our subsequent prediction rule. 

The majority of methods for survival time prediction address censoring using partial likelihood methods, which use event orderings rather than the times at which they occur. Consequently, when survival time can be predicted with reasonable accuracy, partial likelihood methods may miss useful information in the censoring times. Alternatively, some methods use a two-step approach to first impute censored survival times (e.g., mean, median, or multiple imputation), and then fit a predictive model using the imputed survival times \citep{datta2007predicting,wu2008method}. 

Some other methods iteratively impute the censored survival times and fit a predictive model, for example, using survival trees \citep{zhu2012recursively} or an ensemble model \citep{deng2016predicting}. 
%\citet{zhu2012recursively} propose to (i) fit a survival tree to the censored training data, (ii) generate new datasets by imputing the censored observations conditional on the fitted model and censoring times, and (iii) refit the survival tree to the imputed datasets. Similarly, in the 2015 Prostate Cancer Dream Challenge, the winning method, described in \citet{deng2016predicting}, used an algorithm that iterated between (i) fitting an ensemble model to an imputed dataset and (ii) updating the imputed survival times conditional on the fitted model from (i). 
% -- Cut for space: \citet{pan2000multiple} developed an approach for interval and right censored data which iteratively imputes the interval censored outcomes (but not right censored outcomes) and fits a Cox model to the imputed data. 
These methods were not designed for ultra-high dimensional -omic data. For example, in their real data analysis examples, the sample size ($n$) and the number of covariates ($p$) are 
% -- removed after cutting Pan et al reference: $n=94$ and $p=1$ for \citet{pan2000multiple}, 
$n=686$ and $p=8$ for \citet{zhu2012recursively}, and $n=2070$ and $p=256$ for \citet{deng2016predicting}. In contrast, we propose a new method that iteratively imputes the censored survival times and fits a kernel-based predictive model. Our real data analysis has much higher dimensionality than the earlier methods with $n=513$ and $p=20,428$. \cite{zhu2017integrating} also employed a kernel-based method for survival time prediction using gene expression, though they only used one kernel derived from gene expression and did not seek to impute the censored survival times. 
%Our method naturally incorporates high dimensional omics data and is model-based, and thus, easily understood by practitioners. 

The remainder of this article is organized as follows: in Section 2 we described our proposed model and discuss its relation to existing methods; in Section 3 we describe how to compute our estimator; in Section 4 we perform simulation studies to demonstrate our method's prediction accuracy under a range of models; in Section 5 we analyze the TCGA dataset which motivated our study; and in Section 6 we discuss limitations and extensions of our method.

\section{Gaussian process accelerated failure time model}
Let $S_i$ denote the time-to-failure (survival time) for the $i$th patient with $i=1,\dots, n$ patients in the study. Let $T = (\log S_1, \dots, \log S_n)' \in \mathbb{R}^n$. Let $x_i \in \mathbb{R}^p$ and $z_i \in \mathbb{R}^{q+1}$ denote the measured genome-wide gene expression and the measured clinical variables for the $i$th patient, respectively. To allow for an intercept, assume that the first entry of $z_{i}$ is equal to one for $i = 1, \dots, n$. Let $Z = (z_1, \dots, z_n)' \in \mathbb{R}^{n \times (q+1)}$, and $X = (x_1, \dots, x_n)' \in \mathbb{R}^{n \times p}$. For the $n$ patients in the study, we assume that survival time follows the \textit{Gaussian process accelerated failure time model}: 
\begin{equation} \label{eq:accelerated_failure_time}
T = Z \bbeta + G + \epsilon, \quad G \sim {\rm N}_n\left\{ 0, K(X, \bsigma^2) \right\}, \quad \epsilon \sim {\rm N}_{n}\left\{0, \sigma_{\epsilon}^2 I_n\right\},
\end{equation}
where $G$ and $\epsilon$ are independent; $\bsigma^2 \in \mathbb{R}^M_+, \sigma_{\epsilon}^{2} \in \mathbb{R}_+$, and $\bbeta\in\mathbb{R}^{q + 1}$ are unknown model parameters, $\mathbb{R}_+$ denotes non-negative real numbers, and $M$ is the number of kernels. We will sometimes use the more compact notation: 
${\rm Cov}(G + \epsilon) \equiv \tilde{K}(X, \tilde{\bsigma}^2) = K(X, \bsigma^2) + \sigma_{\epsilon}^2 I_n,$
where $\tilde{\bsigma}^2 = ({\bsigma^{2}}', \sigma^2_\epsilon)' \in \mathbb{R}^{M + 1}_{+}.$

The function $K: \mathbb{R}^{n \times p} \times \mathbb{R}^M_+ \to \mathbb{S}^{n}_+$ is a covariance function with $(i,j)$th entry $$[K(X, \bsigma^2)]_{i,j} = \sum_{s=1}^M \sigma_s^2 k_s(x_i, x_j), \quad (i,j) \in \left\{ 1, \dots,n \right\} \times \left\{1, \dots,n \right\},$$ 

where $\mathbb{S}^{n}_+$ denotes the set of $n \times n$ symmetric and positive definite matrices, and $k_s:\mathbb{R}^p \times \mathbb{R}^p \to \mathbb{R}$ is a positive definite kernel function for $s = 1, \dots, M$. A positive definite kernel function ensures that the matrix $k_s(X, X)$: $\mathbb{R}^{n \times p} \times \mathbb{R}^{n \times p} \to \mathbb{S}^{n}_+$, whose $(i,j)$th entry is $k_s(x_i, x_j)$, is positive definite for all $X \in \mathbb{R}^{n \times p}$. The function $k_s(x_i, x_j)$ quantifies the similarity between $x_i$ and $x_j$, e.g., a radial basis kernel function is $k_s(x_i, x_j) = {\rm exp}(-\|x_i - x_j\|^2)$. 

The Gaussian process accelerated failure time model in \eqref{eq:accelerated_failure_time} generalizes the log-normal accelerated failure time model of \citet{klein1999modeling}, which for clustered subjects, assumed that ${\rm Cov}(T_i, T_j) = \phi$ for all $(i,j)$ such that $i$ and $j$ belong to the same cluster and $i \neq j$. Gaussian processes have also been used for survival analysis under the Cox proportional hazards model \citep{banerjee2003frailty, fernandez2016gaussian,zhu2017integrating}. 
Intuitively, \eqref{eq:accelerated_failure_time} assumes that if two patients have similar genome-wide gene expression, as defined by $K$, then their mean-adjusted log-survival times will be similar. Out-of-sample prediction based on \eqref{eq:accelerated_failure_time} is also known as \textit{kriging}, a method for prediction through linear interpolation in geo-spatial statistics. In geo-spatial applications, the function $K$ is used to quantify the similarities of two-dimensional coordinates, whereas in our application, $K$ quantifies similarities in an ultra-high dimensional,  genome-wide space. Recently, kriging was applied in the genomic literature as a means for predicting a phenotypic trait using multiple types of -omics data \citep{wheeler2014poly}. 

Fitting \eqref{eq:accelerated_failure_time} is non-trivial when one observes a censored realization of $T$, as is often the case in survival analysis. Specifically, suppose there exists a realization $T =  (t_1, \dots, t_n)' \in \mathbb{R}^n,$ which cannot be observed. Instead one observes the pairs $(y_1, \delta_1), \dots, (y_n, \delta_n)$ where 
$ y_i = \min(t_i, d_i)$, $d_i$ is the censoring time for the $i$th subject, $\delta_i  = 
1(y_i = t_i)$ for $i = 1, \dots, n$, and $1(\cdot)$ is an indicator function.  In this article, we treat the censored survival times as missing. This allows us to develop an algorithm that simultaneously imputes the latent survival times conditional on the observed survival times and model parameters; and estimates model parameters $\bbeta, \bsigma^2, \sigma^2_{\epsilon}$. 
Although we focus on the case of right-censored outcomes, our methodology naturally accommodates right, left, and interval censoring.

For the remainder of the article, without loss of generality, suppose that $\delta_i = 0$ for $i=1, \dots, n_c$, $\delta_i = 1$ for $i = n_c + 1, \dots, n$, and let $n_o = n - n_c$. Hence, we can partition $Y = (y_1, \dots, y_n)'$ into $Y_{\rm c}\in \mathbb{R}^{n_{\rm c}}$ and $Y_{\rm o} \in \mathbb{R}^{n_{\rm o}}$ so that $Y = (Y_c', Y_o')' \in \mathbb{R}^n$.  We similarly partition $T$ into $(T_c', T_o')'$ (where $T_c$ is not observed and $T_o = Y_o$);  $Z$ into $Z_c \in \mathbb{R}^{n_c \times (q+1)}$ and $Z_o \in \mathbb{R}^{n_o \times (q+1)}$; and $\tilde{K}(X, \tilde{\bsigma}^2)$ into sub-matrices $\tilde{K}_{co}(X, \tilde{\bsigma}^2) \in \mathbb{R}^{n_c \times n_o}$, $\tilde{K}_{oo}(X, \tilde{\bsigma}^2) \in \mathbb{R}^{n_o \times n_o}$, and  $\tilde{K}_{cc}(X, \tilde{\bsigma}^2) \in \mathbb{R}^{n_c \times n_c}$. For ease of display, we will sometimes omit the $(X, \tilde{\bsigma}^2)$ dependence on $\tilde{K}(X, \tilde{\bsigma}^2)$ and its submatrices. Let $\mathcal{H} =  \mathbb{R}^{q + 1} \times \mathbb{R}_+^{M} \times \mathbb{R}_+$ denote the space of the unknown parameters $\theta = (\bbeta', \bsigma^2, \sigma^2_{\epsilon})'.$  Finally, let $W$ be the collection of data that we condition on: $W = \left\{Z, X, Y, \delta\right\}$.

\section{Maximum likelihood estimation}
\subsection{Overview}
To fit the Gaussian process accelerated failure time model in \eqref{eq:accelerated_failure_time}, we use a Monte Carlo expectation-maximization (MC-EM) algorithm. We provide an overview of the MC-EM algorithm in Section 3.2 and describe the sub-algorithms used for distinct covariance function specifications in Section 3.3. We implement our MC-EM algorithm, along with a set of auxiliary functions, in ab R package \texttt{SurvGPR}, which is available in the Supplementary Materials. 

\subsection{Monte Carlo expectation-maximization algorithm}
Throughout this section, let the superscript $(r)$ denote the $r$th iterate of the MC-EM algorithm, and let $s_r$ denote the $r$th iterate's Monte Carlo sample size. 

The $(r+1)$th iterate of the standard expectation-maximization (EM) algorithm is computed in two steps: the E-step computes
\begin{equation}\label{eq:E_step}
Q(\theta \mid \theta^{(r)}) = {\rm E} \left[ \log f_{T} (T_o, T_c; \theta, W) \mid \theta^{(r)}, W\right],
\end{equation}
where $\log f_{T}$ is the log-likelihood of $T$; and the M-step computes 
\begin{equation}\label{eq:Theta} 
\theta^{(r+1)} = \argmax_{\theta \in  \mathcal{H}} Q(\theta \mid \theta^{(r)}).
\end{equation} 
When \eqref{eq:Theta} cannot be obtained, an alternative is to compute $\theta^{(r+1)}$ such that 
\begin{equation}\label{eq:gem_iterate}
\theta^{(r+1)} \in \left\{\theta \in \mathcal{H}: Q(\theta \mid \theta^{(r)}) \geq Q( \theta^{(r)} \mid \theta^{(r)}) \right\}, 
\end{equation}
which yields the generalized EM algorithm \citep{wu1983convergence}.

Unfortunately, when log-survival times are censored, there may not exist an analytic expression for the right hand side of \eqref{eq:E_step} under \eqref{eq:accelerated_failure_time}. In particular, ignoring constants, 
\begin{align}
Q(\theta \mid \theta^{(r)}{}) & \propto - {\rm E}\left[\log {\rm det} \{\tilde{K} (X, \tilde{\bsigma}^2)\} + (T - Z\bbeta)'\{ \tilde{K}(X, \tilde{\bsigma}^2)\}^{-1} (T - Z\bbeta) \mid \theta^{(r)}, W\right] \label{E_step},
\end{align}
so that computing $Q(\theta \mid \theta^{k})$ requires evaluating
$$
(i) \hspace{2pt} {\rm E}\left[ T_{\rm c}\mid \theta^{(r)}, W\right], \quad  (ii) \hspace{2pt} {\rm E}\left[ T_{\rm c}'(\tilde{K}_{cc} - \tilde{K}_{co}\tilde{K}_{oo}^{-1}\tilde{K}_{co}')^{-1} T_{\rm c}\mid  \theta^{(r)}, W \right].
$$
Computing $(i)$ and $(ii)$ is non-trivial because 
\begin{equation} \label{Tc_Dist}
 T_c \mid \theta^{(r)}, W \sim {\rm N}^{[Y_c, \infty)}_{n_c} \left\{ Z_c \bbeta^{(r)} + \tilde{K}_{co}\tilde{K}_{oo}^{-1}(T_o - Z_o \bbeta^{(r)}),  \tilde{K}_{cc} - \tilde{K}_{co}\tilde{K}_{oo}^{-1}\tilde{K}_{co}' \right\}, 
\end{equation}
where the notation ${\rm N}_{n_c}^{[Y_c, \infty)}$ denotes the $n_c$-dimensional truncated multivariate normal distribution with nonzero probability mass on the hyper-rectangle $[Y_c, \infty)=  [y_1, \infty) \times \dots \times [y_{n_c}, \infty).$
Although $(i)$ can be computed numerically, the distribution of $(\tilde{K}_{cc} - \tilde{K}_{co}\tilde{K}_{oo}^{-1}\tilde{K}_{co}')^{-1/2} {T}_c \mid (\theta^{(r)}, W)$ is not truncated multivariate normal unless $(\tilde{K}_{c} - \tilde{K}_{co}\tilde{K}_{oo}^{-1}\tilde{K}_{co}') = I_{n_c}$ \citep{horrace2005some}, so $(ii)$ is intractable in general. Instead, we approximate \eqref{eq:E_step} by drawing $s_r$ samples from \eqref{Tc_Dist} \citep{wei1990monte}.

There are multiple software packages available to simulate from \eqref{Tc_Dist}. In our implementation, we use the Gibbs sampler implemented in the \texttt{tmvtnorm} package in R \citep{tmvtnorm}.  Let $\tilde{T}_{c}^{(r)} = (T_{c,1}^{(r)}, \dots, T_{c,s_r}^{(r)})' \in \mathbb{R}^{s_r \times n_c}$ be the matrix of samples from \eqref{Tc_Dist}. Given $\tilde{T}_{c}^{(r)},$ the $(r+1)$th iterate of our MC-EM algorithm is
\begin{equation}\label{eq:theta_update}
\theta^{(r+1)} = \argmax_{\theta \in  \mathcal{H}} \left\{s_r^{-1} \sum_{j=1}^{s_r}\log f_{T} (T_o, T_{c,j}^{(r)}; \theta, W)\right\}.
\end{equation}
We propose an algorithm to compute \eqref{eq:theta_update} in Section 3.3. To improve the efficiency of our MC-EM algorithm, we use the ascent-based variation proposed by \citet{caffo2005ascent}. We state our complete ascent-based MC-EM algorithm in Algorithm 1. 

\begin{itemize}
\item[] \textbf{Algorithm 1:} Initialize $\theta^{(1)} = (\bbeta^{(1)}, \bsigma^{2(1)}, \sigma_{\epsilon}^{2(1)})$. Set $r=1$ and $s_1 = 500$.
\begin{enumerate}
\item Simulate $\tilde{T}_{c}^{(r)}$,  $s_{r}$ samples from
$$ {\rm N}^{[Y_c, \infty)}_{n_c} \left\{ Z_c \bbeta^{(r)} + \tilde{K}_{co}'\tilde{K}_{oo}^{-1}(T_o - Z_o \bbeta^{(r)}),  \tilde{K}_{cc} - \tilde{K}_{co}\tilde{K}_{oo}^{-1}\tilde{K}_{oc} \right\}.$$
\item Compute $\bar{\theta} \leftarrow \argmax_{\theta \in  \mathcal{H}} \left\{s_r^{-1} \sum_{j=1}^{s_r}\log f_{T} (T_o, T_{c,j}^{(r)}; \theta, W)\right\}$.
\item Compute ${\rm ASE}^{(r)}$, the standard error of $ \left\{ \log f_{T} (T_o, T_{c,j}^{(r)}; \bar{\theta}, W) - \log f_{T} (T_o, T_{c,j}^{(r)}; \theta^{(r)}, W)\right\}_{j=1}^{s_r}$.
\item[4a.] If $s_{r}^{-1}\sum_{j=1}^{s_r} \left\{\log f_{T} (T_o, T_{c,j}^{(r)}; \bar{\theta}, W) - \log f_{T} (T_o, T_{c,j}^{(r)}; \theta^{(r)}, W) \right\} > 1.96 {\rm ASE}^{(r)}$
\begin{itemize} 
	\item Set $\theta^{(r+1)} \leftarrow \bar{\theta}$, $s_{r+1} = s_{r}$, $r \leftarrow r + 1$, and return to Step 1. 
\end{itemize}
\item[4b.] Else
\begin{itemize}
\item If $s_r \geq 10^5,$ terminate. Else, set $s_{r} \leftarrow 2 s_{r}$ and return to Step 1, appending $s_r$ new samples to the $s_r$ from the previous iteration.
\end{itemize}
\end{enumerate}
\end{itemize}

We terminate the algorithm based on a Monte Carlo sample size threshold in Step 4. When the algorithm has converged, the difference between $\theta^{(r)}$ and $\bar{\theta}$ will be negligible for sufficiently large $s_r$. In practice, we suggest practitioners track the parameter estimates across iterations to ensure that $10^5$ is a sufficiently large threshold for their application. 

Because we use a Gibbs sampler in Step 1, the simulated $T_{c,j}^{(r)}$ may be correlated. To decrease dependence while maintaining computational efficiency, we keep every tenth sample generated by the Gibbs sampler \citep{owen2017statistically}. To compute the standard error in Step 3 while accounting for the serial correlations due to the Gibbs sampler, we use the spectral variance method with a Tukey-Hanning window implemented in the R package \texttt{mcmcse} \citep{mcmcse}. 

\subsection{Maximization algorithms}
We now describe how to solve Step 2 of Algorithm 1. Throughout this section, treat $r$ as fixed, let $\hat{T}_j = (T_{c,j}^{(r)'}, T_o')$ for $j=1, \dots, s_{r}$, and let $\bar{T} = s_{r}^{-1} \sum_{j=1}^{s_r} \hat{T}_j$. We develop distinct algorithms for solving \eqref{eq:theta_update} for two types of covariance functions: the single kernel case ($M=1$), and the more general case of multiple distinct kernel functions $(M \geq 1)$. For both cases, we solve \eqref{eq:theta_update} using blockwise coordinate descent. The algorithm we use for the case that $M=1$ is described in the Supplementary Material. This algorithm exploits that $k_1(X,X)$ and $\tilde{K}(X, \tilde{\sigma}^2)$ have the same eigenvectors under \eqref{eq:accelerated_failure_time}. 

For the general case that $M \geq 1$, we use a variation of the blockwise coordinate descent algorithm proposed by \citet{zhou2015mm}. The complete algorithm is stated in Algorithm 2. 

\begin{itemize}
\item[]\textbf{Algorithm 2:} Initialize $\theta^{(1)} = (\bbeta^{(1)}, \bsigma^{2(1)}, \sigma_{\epsilon}^{2(1)})$ at their final iterates from the previous M-step. Set $b=1$.  
\vspace{-8pt}
\begin{enumerate}
 \item Compute $\Omega \leftarrow \tilde{K}(X, \tilde{\bsigma}^{2(b)})^{-1}$
  \item Compute $\bbeta^{(b+1)} \leftarrow (Z'\Omega Z)^{-1}Z'\Omega \bar{T}$
  \item For $ i = 1, \dots, M, $ compute
  \vspace{-8pt}
   $$ \sigma_{i}^{2(b+1)} \leftarrow \frac{\sigma_{i}^{2(b)}}{\sqrt{s_r}} \left[ \frac{\sum_{j=1}^{s_{r}}(\hat{T}_j - Z \bbeta^{(b+1)})'\Omega' k_i(X, X) \Omega (\hat{T}_j - Z \bbeta^{(b+1)})}{{\rm tr}\left\{ \Omega k_i(X, X)\right\} } \right]^{1/2},$$
     \vspace{-12pt}

    \item Compute 
      \vspace{-14pt}
   $$ \sigma_{\epsilon}^{2(b+1)} \leftarrow \frac{\sigma_{\epsilon}^{2(b)}}{\sqrt{s_{r}}} \left[ \frac{\sum_{j=1}^{s_r}(\hat{T}_j - Z \bbeta^{(b+1)})'\Omega'\Omega (\hat{T}_j - Z \bbeta^{(b+1)})}{{\rm tr}(\Omega)}\right]^{1/2}.$$
     \vspace{-10pt}
  \item[5a.] If $\{ \sum_{j=1}^{s_r} \log f_{T} (T_o, T_{c,j}^{(r)}; \theta^{(b+1)}, W) - \log f_{T} (T_o, T_{c,j}^{(r)}; \theta^{(b)}, W)\}$\\
  $\leq \epsilon|\sum_{j=1}^{s_r} f_{T} (T_o, T_{c,j}^{(r)}; \theta^{(1)}, W)|$
  \begin{itemize} 
  \item Terminate. 
  \end{itemize}
  \item[5b.] Else
   \begin{itemize} 
  \item Set $b \leftarrow b + 1$ and return to Step 1.
  \end{itemize}
\end{enumerate}
\end{itemize}
The updates of $\sigma_{i}^{2(b+1)}$ and $\sigma_{\epsilon}^{2(b+1)}$ in Steps 3 and 4 are derived based on the minorize-maximize (MM) algorithm for variance components estimation proposed by \citet{zhou2015mm}. Briefly, given the initial values of the parameters or their estimates from the previous iteration, a minorizing function is created to approximate the objective function. The updates in Steps 3 and 4 are the arguments that maximize a minorizing function and thus, ensure that the objective function evaluated at $\theta^{(b+1)}$ is greater than or equal to the objective function evaluated at $\theta^{(b)}.$ A complete derivation of Algorithm 2 is provided in the Supplementary Material. 
% By arguments similar to those in \citet{zhou2015mm}, one can verify that Algorithm 2 has the strict ascent property, i.e., the objective function \eqref{eq:theta_update} evaluated at $\theta^{(b+1)}$ is greater than or equal to the objective function evaluated at $\theta^{(b)}$. Thus, although \eqref{eq:theta_update} is a non-convex optimization problem and we are not guaranteed to obtain its global maximizer, Algorithm 2 ensures that $\sum_{j=1}^{s_r}\log f_{T} (T_o, T_{c,j}^{(r)}; \theta^{(b+1)}, W) \geq \sum_{j=1}^{s_r}\log f_{T} (T_o, T_{c,j}^{(r)}; \theta^{(b)}, W)$. This implies $Q(\theta^{(b+1)} \mid \theta^{(b)} ) \geq Q(\theta^{(b)} \mid \theta^{(b)})$, as in \eqref{eq:gem_iterate}, with high probability for sufficiently large $s_r$.

In our implementation, we also use quasi-Newton-like acceleration attempts based on an extrapolation heuristic. We found that the iterates from Steps 3 and 4 of Algorithm 2 often followed monotonic paths to their local maximizers. Thus, after Step 4, we attempt to replace $\bsigma^{2(b+1)}$ with an extrapolated value 
$$\bar{\bsigma}^{2(b+1)} = \bsigma^{2(b+1)} + (b^{1/2} + 2)^{-1}(\bsigma^{2(b+1)} - \bsigma^{2(b)}),$$ and similarly for $\sigma_{\epsilon}^{2(b+1)}$. If the log-likelihood evaluated at the extrapolated values $\bar{\bsigma}^{2(b+1)}$ and $\bar{\sigma}_{\epsilon}^{2(b+1)}$ is greater than the log-likelihood evaluated at the $\bsigma^{2(b+1)}$ and $\sigma_{\epsilon}^{2(b+1)}$, we replace $\bsigma^{2(b+1)}$ with $\bar{\bsigma}^{2(b+1)}$ and $\sigma_{\epsilon}^{2(b+1)}$ with $\bar{\sigma}_{\epsilon}^{2(b+1)}$.

\subsection{Implementation and practical considerations }
Given the final iterates of the MC-EM algorithm, $\hat{\bbeta},\hat{\bsigma}^2, \hat{\sigma}^2_\epsilon$, and final imputed survival time, $\bar{T}$, we predict log-survival time for a new patient with covariates $z_*$ and genome-wide gene expression $x_*$ using the conditional expectation of the univariate normal distribution:
\begin{equation}\notag
{\rm N}\left\{ \hat{\bbeta}'z_* + K_{*}(x_*, X, \hat{\bsigma}^2)' \tilde{K}(X, \hat{\tilde{\bsigma}}^2)^{-1} (\bar{T} - Z\hat{\bbeta}), \ \  \tilde{K}(x_*, \hat{\tilde{\bsigma}}^2) -  K_{*}(x_*, X, \hat{\bsigma}^2)' \tilde{K}(X, \hat{\tilde{\bsigma}}^2)^{-1}  K_{*}(x_*, X, \hat{\bsigma}^2) \right\},
\end{equation}
where $K_{*}(x_*, X, \hat{\bsigma}^2) \in \mathbb{R}^{n}$ with $j$th entry $[K_{*}(x_*, X, \hat{\sigma}^2)]_j = \sum_{s=1}^M \hat{\sigma}_s^2 k_s(x_*, x_j)$ for $j= 1, \dots, n$. We can also easily evaluate the estimated survival function, $\hat{\mathcal{S}}$ at any time $a$ since $P(T_* < a \mid z_*, x_*)$ is the cumulative distribution function of a univariate normal distribution.

In studies collecting gene expression or other types of -omics data, there are often measured technical confounders, e.g., the plate on which an RNA sample was stored. To address confounding in genome-wide gene expression under \eqref{eq:accelerated_failure_time}, we propose to compute the kernel functions $k_s$ using the residuals from the multivariate regression of gene expression on the measured technical confounders.

To obtain reasonable initial values for our MC-EM algorithm with right-censored survival times, we suggest first imputing the censored log-survival times using the inverse probability weighted mean-imputation method proposed by \citet{datta2005estimating}.

\section{Simulation studies}

\subsection{Data generating models}\label{data_gen_models}
To create simulation scenarios similar to our motivating data example, we  use the observed gene expression data and clinical covariates of the 513 patients in the TCGA KIRC (kidney renal clear cell carcinoma) dataset, and we simulate survival times for these patients. Specifically, we use the observed tumor stage and age as clinical covariates, and use the observed expression of $p=20,483$ genes to generate survival times from four distinct models. More information about how we prepared the TCGA KIRC dataset is given in Section \ref{data_preparation}. For 500 independent replications, we generate $n = 513$ survival times and split the data into a training and testing set of size 413 and 100 respectively. We then fit the model to the censored training data and record the metrics described in Section \ref{sec:metrics}. The data generating models we consider are: 
\begin{enumerate}
	\item[] \textit{Model 1: Gaussian process AFT model.} Log-survival times are generated as a realization of the Gaussian process accelerated failure time model: $$T = Z \bbeta + \eta + \gamma,$$ where 
	$\gamma \sim {\rm N}_n \left\{ 0, 0.5 I_n \right\}$ and $\eta \sim {\rm N}_n \left\{ 0, K(X, \bsigma^2) \right\}$ with $K(X, \bsigma^2)$ defined below and $\bbeta = (6.1, -0.5, -1.2, -2.0, -1\times 10^{-5})$ where the columns of $Z$ corresponds to the intercept, tumor stage II, tumor stage III, tumor stage IV, and age in days. 
	\item[] \textit{Model 2: Normal-Logistic AFT model.} Log-survival times are generated as a realization of the normal-logistic accelerated failure time model, $$T = Z \bbeta + \eta + \kappa,$$ 
	where $\kappa = (\kappa_1, \dots, \kappa_n)'$ with each $\kappa_i$ independent and identically distributed 
	logistic distribution such that $E(\kappa_i) = 0$ and ${\rm Var}(\kappa_i) = 0.5$. Note that logistic distribution has much heavier tails than normal distribution. As in Model 1, $\eta \sim {\rm N}_n \left\{ 0, K(X, \bsigma^2) \right\}$ with $K(X, \bsigma^2)$ defined below; and $Z$ and $\bbeta$ are the same as in Model 1. 
	\item[] \textit{Model 3: Logistic-Logistic AFT model.} Log-survival times are generated as a realization of the logistic-logistic accelerated failure time model, $$T = Z \bbeta + \omega + \kappa,$$ where $\kappa$ is generated in the same manner as in Model 2. To generate $\omega \in \mathbb{R}^n$, we generate $v_1, \dots, v_n$, $n$ independent copies of $V_i \sim {\rm Logistic}$ where $E(V_i) = 0$ and ${\rm Var}(V_i) = 1$ for $i=1, \dots, n$. Then, we set $(\omega_{1}, \dots, \omega_{n})' = ( v_1, \dots, v_n)'\{ K(X, \bsigma^2)\}^{1/2}$ so that ${\rm E}(\omega_i) = 0$ and ${\rm Cov}(\omega_i, \omega_j) = [K(X, \bsigma^2)]_{i,j}.$ 
	\item[] \textit{Model 4: Cox proportional hazards model.} We generate survival times from the mixed-effects Cox proportional hazards model with Gompertz baseline hazard \citep{bender2005generating}. 
	Let $W = \tilde{Z} \bbeta + \eta $ where $\bbeta = (0.1, 0.3, 0.9, 9 \times 10^{-5})$ with columns of $\tilde{Z}$ corresponding to tumor stage II, tumor stage III, tumor stage IV, and age in days, respectively; and let $\eta \sim {\rm N}_n \left\{ 0, K(X, \bsigma^2) \right\}$ with $K(X, \bsigma^2)$ defined below. Then following \cite{bender2005generating}, we generated survival times as a realization of 
	$$ S_i = \frac{1}{\alpha} \log \left[ 1 - \frac{\alpha \log(u_i)}{\lambda {\rm exp}(W_i)} \right], \quad i = 1, \dots, n$$
	where $u_1, \dots, u_n$ are $n$ independent realizations of a $\text{Uniform}(0, 1)$ random variable; and $\alpha = \pi(1200 \sqrt{6})^{-1}$ and $\lambda = \alpha {\rm exp}(-.5772 - \alpha 1400)$ are chosen to mimic the survival distribution in the real dataset with mean $1400$ and standard deviation $1200$. 
\end{enumerate}
For Models 1--4, the $i$th subject's censoring time is drawn from an exponential distribution with mean $\left\{{\rm exp}\left[ Q_{\tau_i}(\{Y_j\}_{j=1}^n)\right] \right\}^{-1}$ where $Q_{c}$ denotes the $c$th quantile and $\tau_i = .20, .50, .70,$ or $.80$ for subjects with tumor stages I, II, III, or IV respectively. Between 62\% - 63\% of survival times are censored on average in each of the four data generating models. 

Models 2 -- 4 illustrate our method's performance under three different types of misspecification. In Model 2, the error distribution is misspecified by our method, whereas in Model 3, both the genomic effect and the error distribution are misspecified by our method. Under Model 4, the log-linearity of survival time is violated. 

\begin{figure}[t]
\centerline{\hfill\includegraphics[width=.50\textwidth]{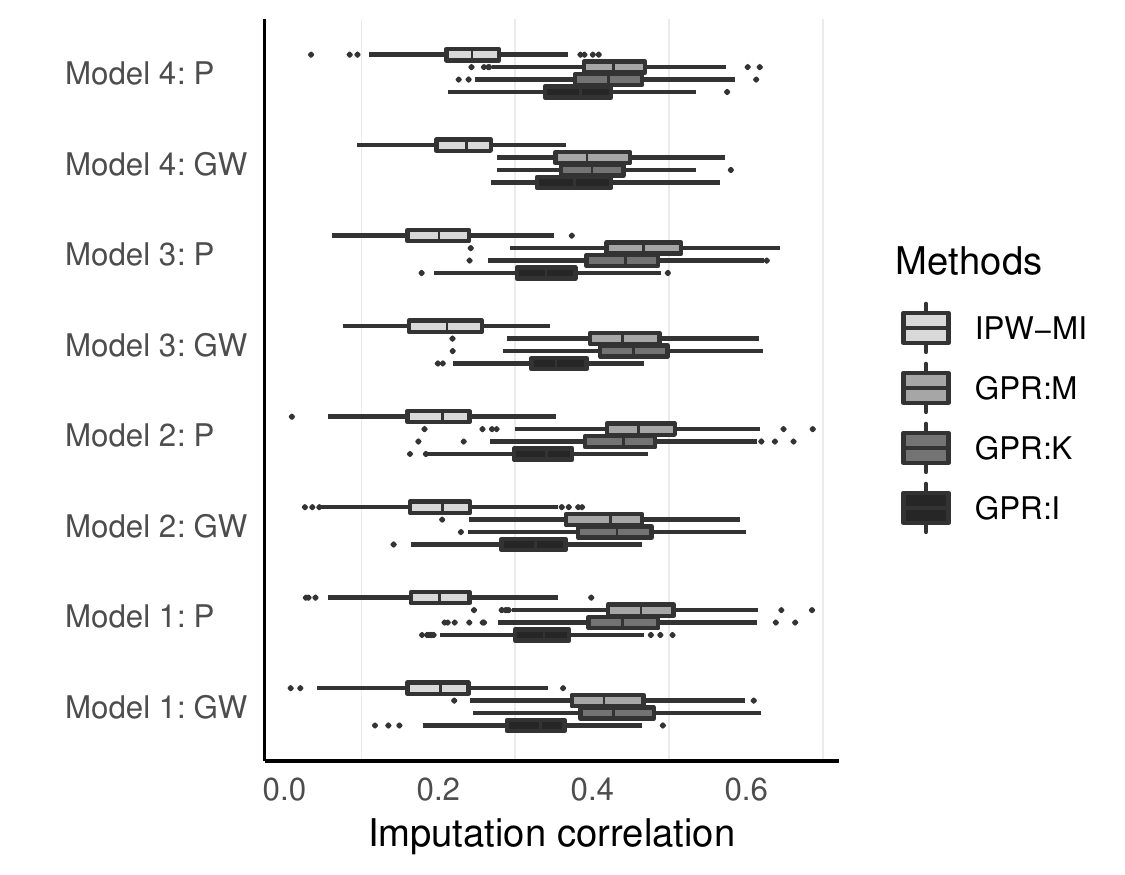}\hfill \includegraphics[width=.50\textwidth]{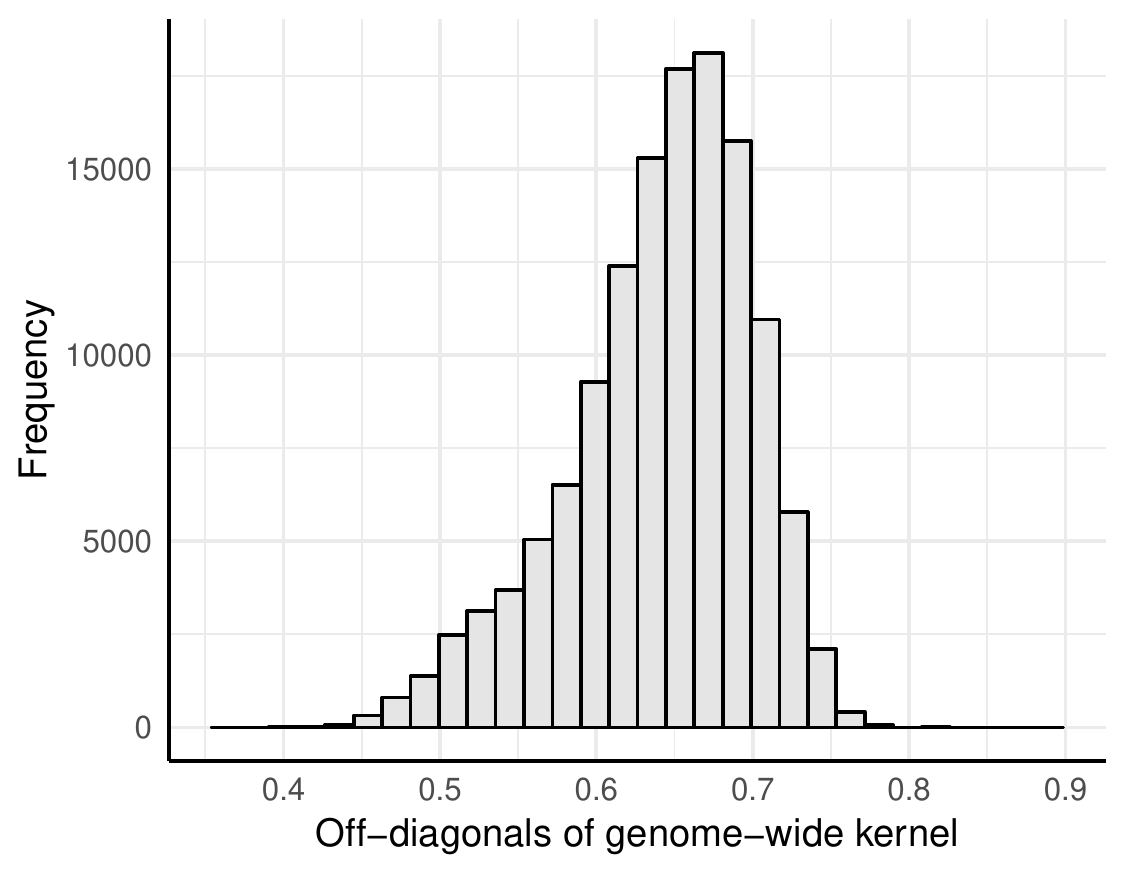}\hfill}
\centerline{\hfill\makebox[.50\textwidth]{(a)}\hfill\makebox[.50\textwidth]{(b)} \hfill}
\caption{(a) Correlations between the imputed log-survival time and true log-survival time for the censored outcomes in the training sets. (b) A histogram showing the off-diagonal entries of the normalized radial basis  kernel using genome-wide gene expression.}\label{fig:aux_information}
\end{figure}

For each of the four data generating models, we consider two variations of covariance function $K(X, \bsigma^2):$
\begin{itemize}
	\item[] {\textit{Genome-wide kernel.}} We compute $K(X, \bsigma^2)$ using a normalized radial basis function kernel based on genome-wide gene-expression. Given $x_i \in \mathbb{R}^p$ for $i=1, \dots, n$, we compute 
	\begin{equation}\label{norm_euc_dist}
	 [K(X, \bsigma^2)]_{i,j} = \sigma_G^2 \underbar{$k$}(x_i, x_j) \equiv \sigma_G^2 \hspace{2pt} {\rm exp}\left\{ \frac{-\|x_i - x_j\|^2}{\max_{l,m} \left(\|x_l - x_m\|^2 \right)}  \right\}.
	 \end{equation}
	 In Models  1 -- 3, we set $\sigma_G^2 = 3$, and in Model 4, we set $\sigma_G^2 = 4.$ A histogram showing the off-diagonal entries of \eqref{norm_euc_dist} is displayed in Figure \ref{fig:aux_information}(b).
	\item[] {\textit{Pathway kernel.}} We compute $K(X, \bsigma^2)$ as the sum of normalized radial basis function kernels as in \eqref{norm_euc_dist} based on a small number of genes meant to emulate gene-pathways: 
	\begin{equation}\label{eq:pathway_kernels}
	[K(X, \bsigma^2)]_{i,j} = \sum_{s=1}^6 \sigma_s^2 \underbar{$k$}(D_s x_i, D_s x_j),\quad (i,j) \in \left\{ 1, \dots, n\right\} \times \left\{ 1, \dots, n\right\},
	\end{equation} where for $s=1, \dots, 6$, $D_s \in \mathbb{R}^{p \times p}$ has 150, 150, 100, 100, 50, and 50 randomly positioned ones on its diagonal and zeros in all other positions. Four of six $\sigma_s^2$ are randomly assigned to be nonzero and their magnitudes are drawn independently ${\rm Uniform}[0,1]$ and normalized so that $\sum_{s=1}^6 \sigma_s^2 = 3$ in Models 1 -- 3, and  $\sum_{s=1}^6 \sigma_s^2 = 4$ in Model 4. 
\end{itemize}
The genome-wide kernel data generating model illustrates the performance of the competing methods when effect sizes are small and abundant, i.e., genome-wide. The pathway-kernel data generating model is meant to compare our method to those which perform variable selection or marginal screening since there will be at most five hundred genes that affect the survival time distribution. 

\subsection{Methods}
In their review of methods for predicting survival time based on gene expression, \citet{van2009survival} concluded that among the tree-based ensemble methods and regularized Cox proportional hazard models they compared, the $L_2$-penalized Cox proportional hazards model performed as well or better than the other methods. For this reason, we exclude tree-based ensemble methods from our comparisons, but include the $L_1$ and $L_2$-penalized Cox-proportional hazards model using genome-wide gene expression, tumor stage, and age as covariates. We do not penalize coefficients corresponding to tumor stage or age in either model and select tuning parameters using ten-fold cross-validation. We also consider $L_1$ and $L_2$-penalized Cox proportional hazards models using a pre-screened  gene sets, which we call $L_1^{*}$ and $L_2^{*}$. When the data generating model uses the genome-wide kernel, the screening method retains genes that are significantly associated with survival with false discovery rate below 0.10. When the data generating model uses the pathway kernels, the screening method retains the six-hundred genes that are used to construct the pathway kernels, i.e., assuming an oracle screening method. 

We use two variations of our Gaussian process accelerated failure time model. The versions denoted \texttt{GPR:K} and \texttt{GPR:M} correspond to the genome-wide kernel and pathway kernels, respectively. For \texttt{GPR:K}, we use \eqref{norm_euc_dist} as the lone candidate kernel. For \texttt{GPR:M}, we use seven candidate kernels: the six pathway kernels from \eqref{eq:pathway_kernels}, and a genome-wide kernel which is similar to \eqref{norm_euc_dist}, except it uses all genes not used in any of the pathway kernels. When the data-generating model uses the genome-wide kernel, the six pathway kernels are computed from 600 randomly selected genes. When the data generating model uses the pathway-kernels, \texttt{GPR:M} includes the true pathway-kernels as candidates. 

To illustrate the benefit of our imputation procedure, which jointly imputes censored survival times and estimates model parameters, we also compare our method to two versions of the Gaussian process accelerated failure time model that imputes survival times using the inverse probability weighted mean imputation procedure of \citet{datta2005estimating}. These approaches, which we call \texttt{LMM:K} and \texttt{LMM:M}, fit variations of \eqref{eq:accelerated_failure_time} using the MM-algorithm from \citet{zhou2015mm}, treating the imputed survival times as fixed. 

Finally, we also consider two variations of the mixed-effects Cox proportional hazards model used in \citet{zhu2017integrating}. These versions are denoted \texttt{ME-Cox:K} and \texttt{ME-Cox:M} and use the same candidate kernels as \texttt{GPR:K} and \texttt{GPR:M}, respectively.

\subsection{Performance metrics}\label{sec:metrics}
As noted in \citet{van2009survival}, there is no consensus on which metric to use for evaluating the accuracy of prediction in survival analysis. For this reason, we use three different metrics. The first metric is based on the C-index measurement proposed by \citet{uno2011c}. Given $\hat{S}$, the $n_v$ predicted survival times or risk scores for the testing set, we define the C-index as:
$$\frac{\sum_{i=1}^{n_v} \sum_{j=1}^{n_v} \delta_i \{\hat{H}(S_i)\}^{-2} 1(S_i < S_j, S_i < \tau) 1( \mathcal{C}(\hat{S_i}) > \mathcal{C}(\hat{S_j}))}{\sum_{i=1}^{n_v} \sum_{j=1}^{n_v} \delta_i \{\hat{H}(S_i)\}^{-2} 1(S_i < S_j, S_i < \tau)},$$
where $\mathcal{C}:\mathbb{R} \to \mathbb{R}$ maps a predicted survival time to the risk score scale, $1(\cdot)$ is the indicator function, $\tau$ is the study length, and $\hat{H}(\cdot)$ is the Kaplan-Meier estimator of the censoring distribution. 

The second metric we use is the integrated Brier score. While C-index evaluates the prediction of accuracy in terms of relative order of survival times, the Brier score quantifies the accuracy of survival time function estimate at time $t$:
$$ B(t) = {n_v}^{-1} \sum_{i=1}^{n_v} \left\{ [\hat{\mathcal{S}}(t|z_i, x_i)]^2 \min \left\{ 1(S_i \leq t), \delta_i \right\} \{ \hat{H}(S_i)\}^{-1} + [1 - \hat{\mathcal{S}}(t|z_i, x_i)]^2 1(S_i > t) \{ \hat{H}(S_i)\}^{-1} \right\},$$
where $\hat{\mathcal{S}}(t|z_i, x_i)$ is the estimated survival function for the $i$th subject in the testing data evaluated at time $t$. The integrated Brier score we use is
$$ {\rm BS} = \tau^{-1} \int_{0}^{\tau} B(t) {\rm d} t.$$

The third metric we use is integrated AUC based on the sensitivity and specificity measures defined by \citet{uno2007evaluating}. This AUC-based metric measures the accuracy of $t$-year survival prediction. We use the function \texttt{AUC.uno} from the R package \texttt{SurvAUC} in our implementation. 

In the simulation study, we have access to the true survival times for the entire testing set, so $\hat{H}(Y_j) = 1$ and $\delta_j = 1$ for all $j$ in the testing set. Thus, in the simulation study, we set $\tau$ equal to the largest survival time in the testing data. In the real data analyses, we estimate $\hat{H}$ using the Kaplan-Meier estimator and set $\tau$ equal to the second largest observed survival time in the testing data. 

\subsection{Results}
Simulation results for 500 independent replications are displayed in Table 1. To compare methods, for each replication we record the ratio of each method's performance to the best performance amongst all methods. For C-index and integrated-AUC, a ratio less than one indicates worse performance, whereas for integrated Brier score, a ratio great than one indicates worse performance.  
 
 \begin{table}[t!]
\centering
\scalebox{.78}{
\setlength\tabcolsep{4.5pt} % default value: 6pt
\begin{tabular}{|c|c|c|cccc|cc|cc|cc|c|}
  \hline
 & \multirow{2}{*}{Covariance} & \multirow{2}{*}{Metric} &  \multicolumn{4}{c|}{Regularized Cox} & \multicolumn{2}{c|}{ME Cox} & \multicolumn{2}{c|}{GPR} & \multicolumn{2}{c|}{LMM} & \multirow{2}{*}{Scale}\\
 & & &  $L_1$ & $L_1^*$ & $L_2$ & $L_2^*$ & $K$ & $M$ & $K$ & $M$ & $K$ & $M$ & \\ 
 \hline
 \hline
 \multirow{6}{*}{Model 1}  & \multirow{3}{*}{Genome-wide} &  Brier & 1.175 & 1.166 & 1.153 & 1.148 & 1.180 & 1.165 & \cellcolor{gray!70}1.012 & 1.021 & 1.140 & 1.140 & 0.050 \\ 
&&  AUC & 0.964 & 0.964 & 0.983 & 0.984 & 0.945 & 0.943 & \cellcolor{gray!70}0.988 & 0.983 & 0.975 & 0.969 & 0.796 \\ 
 && C-index & 0.971 & 0.971 & 0.979 & 0.980 & 0.954 & 0.953 & \cellcolor{gray!70}0.992 & 0.988 & 0.980 & 0.975 & 0.709 \\ 
   \cline{2-14}
  &  \multirow{3}{*}{Pathway} & Brier & 1.177 & 1.173 & 1.157 & 1.162 & 1.186 & 1.169 & 1.028 & \cellcolor{gray!70}1.015 & 1.164 & 1.156 & 0.050 \\ 
 && AUC & 0.953 & 0.959 & 0.972 & 0.978 & 0.931 & 0.936 & 0.977 & \cellcolor{gray!70}0.990 & 0.963 & 0.976 & 0.804 \\ 
 && C-index & 0.960 & 0.965 & 0.968 & 0.973 & 0.940 & 0.945 & 0.980 & \cellcolor{gray!70}0.992 & 0.968 & 0.979 & 0.716 \\ 
\hline\hline
\multirow{6}{*}{Model 2}  & \multirow{3}{*}{Genome-wide} & Brier & 1.168 & 1.156 & 1.145 & 1.137 & 1.169 & 1.155 & \cellcolor{gray!70}1.013 & 1.021 & 1.133 & 1.134 & 0.048 \\ 
&& AUC & 0.961 & 0.961 & 0.984 & 0.982 & 0.946 & 0.945 & \cellcolor{gray!30}0.986 & 0.981 & 0.972 & 0.966 & 0.793 \\ 
 && C-index & 0.968 & 0.968 & 0.979 & 0.979 & 0.955 & 0.955 & \cellcolor{gray!70}0.990 & 0.985 & 0.978 & 0.972 & 0.708 \\ 
   \cline{2-14}
  &  \multirow{3}{*}{Pathway} &Brier & 1.179 & 1.177 & 1.160 & 1.165 & 1.183 & 1.169 & 1.030 & \cellcolor{gray!70}1.016 & 1.161 & 1.153 & 0.049 \\ 
 && AUC & 0.954 & 0.958 & 0.974 & 0.978 & 0.933 & 0.938 & 0.976 & \cellcolor{gray!70}0.990 & 0.962 & 0.974 & 0.807 \\ 
 && C-index & 0.960 & 0.964 & 0.969 & 0.973 & 0.942 & 0.946 & 0.980 & \cellcolor{gray!70}0.992 & 0.967 & 0.978 & 0.719 \\ 
\hline\hline
\multirow{6}{*}{Model 3}  & \multirow{3}{*}{Genome-wide}& Brier & 1.179 & 1.168 & 1.158 & 1.150 & 1.180 & 1.165 & \cellcolor{gray!70}1.020 & 1.030 & 1.127 & 1.124 & 0.046 \\
 && AUC & 0.963 & 0.964 & 0.984 & 0.984 & 0.942 & 0.942 & \cellcolor{gray!70}0.988 & 0.982 & 0.966 & 0.973 & 0.794 \\
&&  C-index & 0.971 & 0.971 & 0.981 & 0.981 & 0.952 & 0.951 & \cellcolor{gray!70}0.991 & 0.986 & 0.972 & 0.978 & 0.711 \\
   \cline{2-14}
  &  \multirow{3}{*}{Pathway} & Brier & 1.186 & 1.184 & 1.166 & 1.172 & 1.199 & 1.181 & 1.028 & \cellcolor{gray!70}1.019 & 1.149 & 1.142 & 0.045 \\
  && AUC & 0.957 & 0.961 & 0.975 & 0.980 & 0.935 & 0.938 & 0.980 & \cellcolor{gray!70}0.991 & 0.966 & 0.975 & 0.805 \\
 && C-index & 0.963 & 0.966 & 0.970 & 0.974 & 0.943 & 0.946 & 0.983 & \cellcolor{gray!70}0.992 & 0.970 & 0.977 & 0.720 \\
\hline\hline
\multirow{6}{*}{Model 4}  & \multirow{3}{*}{Genome-wide} &  Brier & 1.056 & 1.065 & \cellcolor{gray!30}1.043 & 1.052 & 1.068 & 1.060 & 1.150 & 1.157 & 1.077 & 1.079 & 0.093 \\ 
 && AUC & 0.971 & 0.968 & \cellcolor{gray!70}0.988 & 0.983 & 0.961 & 0.961 & 0.980 & 0.977 & 0.972 & 0.967 & 0.754 \\ 
 && C-index & 0.979 & 0.977 & 0.985 & 0.982 & 0.970 & 0.971 & \cellcolor{gray!30}0.986 & 0.984 & 0.979 & 0.975 & 0.678 \\ 
   \cline{2-14}
  &  \multirow{3}{*}{Pathway} &  Brier & 1.064 & 1.062 & \cellcolor{gray!30}1.049 & 1.050 & 1.077 & 1.069 & 1.145 & 1.142 & 1.091 & 1.082 & 0.092 \\ 
 && AUC & 0.966 & 0.967 & 0.983 & \cellcolor{gray!20}0.986 & 0.955 & 0.957 & 0.975 & 0.980 & 0.965 & 0.968 & 0.759 \\ 
 && C-index & 0.974 & 0.976 & 0.981 & 0.983 & 0.966 & 0.967 & 0.981 & \cellcolor{gray!70}0.987 & 0.974 & 0.976 & 0.682 \\ 
\hline
\end{tabular}
}
\caption{Average relative performance for five hundred independent replications under the four models. Relative performance is defined as the ratio of each method's error to the best amongst all the competing methods, so that a relative performance of one indicates that the method performed best amongst all the methods. For integrated AUC and C-index, a relative error less than one indicates worse performance, whereas for integrated Brier score, a relative error of greater than one indicates worse performance. Cells highlighted in dark grey indicate a relative performance significantly better than all other methods. Cells highlighted in light grey indicate the best average relative performance, but one not significantly better than all other methods. The scale column displays the mean of the best method's metric across the five hundred replications.}
\end{table}

Under Models 1 -- 3, our proposed methods \texttt{GPR:K} and \texttt{GPR:M} were the best amongst the competing methods in terms of integrated Brier score. For integrated AUC and C-index, the version of our method with correctly specific covariance was best. Under Model 4, our method performs poorly relative to the Cox proportional hazards models and the IPW-imputed version of our method in terms of integrated Brier score and AUC, but performs nearly as well or better than the Cox  proportional hazards models in terms of C-index. This may imply that the Cox model could provide more accurate estimates of the survival function, but less accurate prediction of relative order. 

It is also important to analyze the performance of \texttt{GPR:K} when the covariance $K(X, \bsigma^2)$ is generated from the data generating models using pathway kernels. In general, \texttt{GPR:K}, which uses genome-wide gene expression, performs only slightly worse than \texttt{GPR:M} in terms of C-index and integrated Brier score, and is often similar to the $L_2$-penalized Cox proportional hazards model with screening in terms of integrated AUC. Thus, even though only approximately 2.5\% of genes actually contribute to the genomic effect under the pathway-based data generating model, using genome-wide gene-expression does not seem to degrade prediction accuracy drastically.  

In Figure 2a, we display boxplots showing the correlation between the true log survival time and the imputed log survival for the censored training data. The method \texttt{GPR:I}, which fits \eqref{eq:accelerated_failure_time} assuming $G=0$, performs worse than either \texttt{GPR:K} or \texttt{GPR:M}, all of which perform better substantially better than the IPW mean imputation procedure of \citet{datta2005estimating}, denoted \texttt{IPW-MI}. This partly explains the relative advantage of \texttt{GPR} over \texttt{LMM}. 

\section{KIRC data analysis}
\subsection{Data preparation}\label{data_preparation}
We downloaded demographic/clinical data as well as RNA-seq data (from workflow HTSeq - Counts) of TCGA KIRC patients from NCI  Genomic Data Commons (\href{http://portal.gdc.cancer.gov/}{http://portal.gdc.cancer.gov/}). Data were pre-processed in the following steps.  We omitted patients for whom gene expression was measured on a plate with fewer than ten patients and patients who did not have a measured tumor stage. To filter out genes with relatively low expression in most samples, we use genes whose 75\% percentile read-count was greater than 20. For the 20,428 genes with sufficient read counts, the expression measurement $x_{ik}$, i.e., the expression for the $i$th individual and $k$th gene, is $\log_{10}[(t_{ik} + 1)/q_{i, 0.75}]$ where $q_{i,0.75}$ is the 75th percentile of read counts for the $i$th individual and $t_{ik}$ is the read count for the $i$th individual's $k$th gene. For the resulting dataset consisting of 513 patients with gene expression measured on 20,428 genes, we found that tumor stage, age, and sequencing plate had significant marginal associations with survival under the Cox proportional hazards model. Following \citet{zhu2017integrating}, we also include gender as a covariate.

\begin{figure}[t]
\centerline{\hfill\includegraphics[width=.45\textwidth]{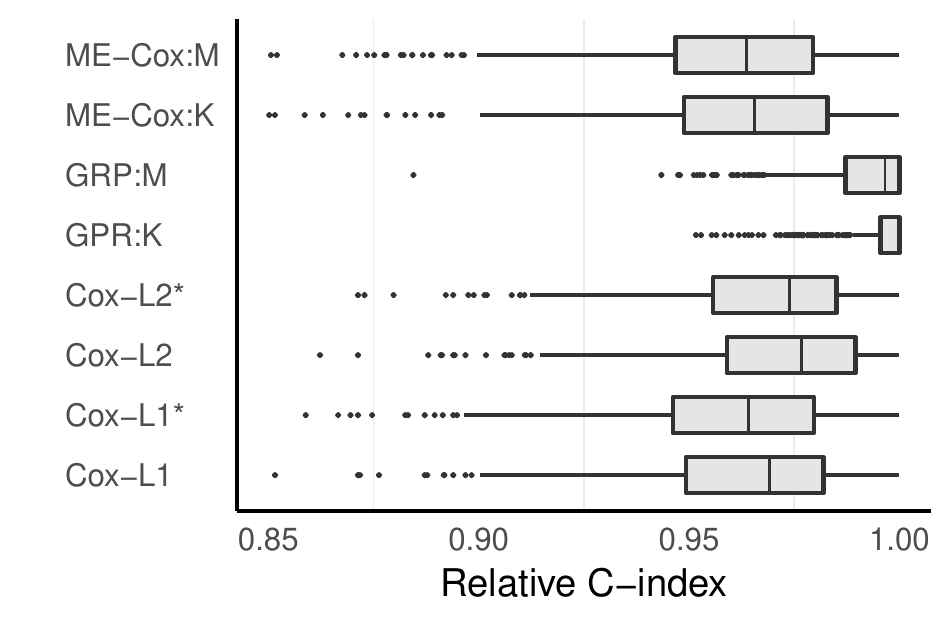}\hfill\includegraphics[width=.45\textwidth]{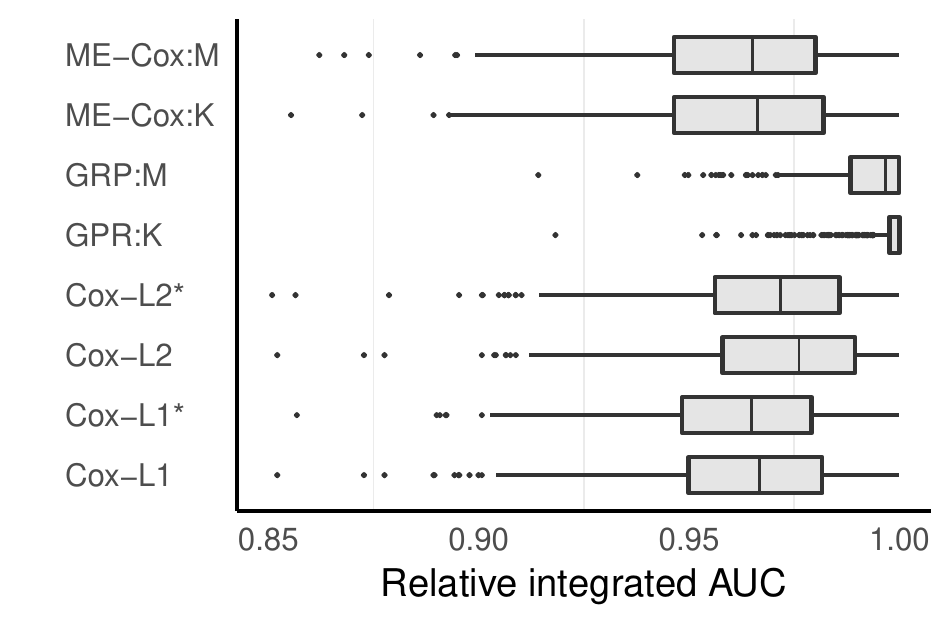}\hfill}
\centerline{\hfill\makebox[2.5cm]{(a) }\hfill \makebox[2.5cm]{\hspace{100pt}(b)} \hfill}
\centerline{\hfill\includegraphics[width=.45\textwidth]{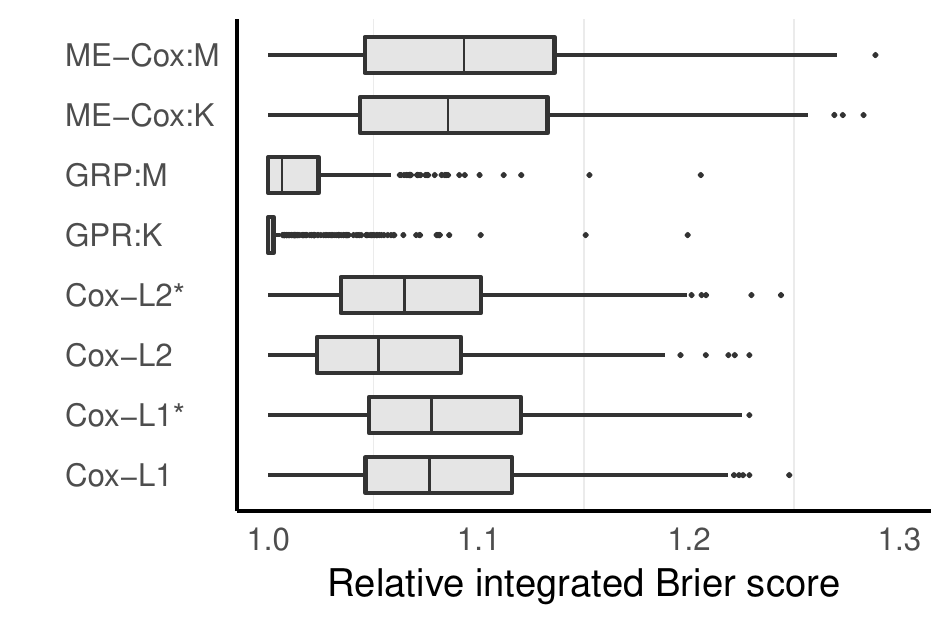}\hfill\includegraphics[width=.45\textwidth]{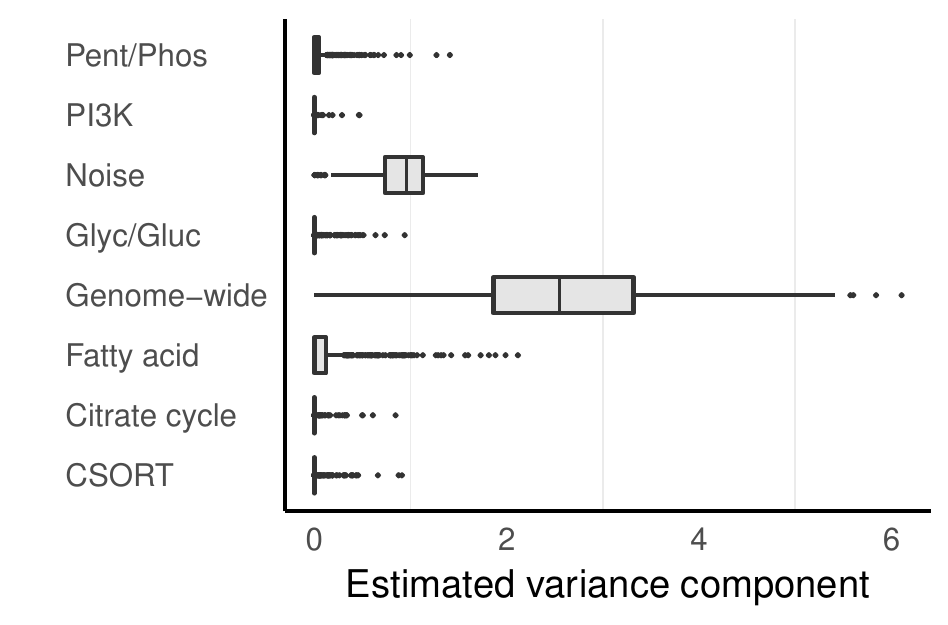}\hfill}
\centerline{\hfill\makebox[2.5cm]{(c) }\hfill \makebox[2.5cm]{\hspace{100pt}(d)} \hfill}
\caption{(a) - (c) Results from 500 independent training/testing splits. Each point represents the method's performance relative to the best performing method within one replication. For (a) and (b), a relative C-index or relative integrated AUC less than one indicates worse performance. For (c), a relative Brier score of greater than one indicates a worse performance. (d) Estimated variance components for each of the candidate kernel in \texttt{GPR:M}. }\label{fig:Analysis_results}
\end{figure}

\subsection{Pathway-based candidate kernels}
Following our simulation studies, we propose to use the normalized radial basis kernel $\underbar{$k$}(\cdot, \cdot)$ to define the kernel function $K$. 
We consider multiple variations of our method: the genome-wide version, \texttt{GPR:K}, which uses the normalized radial basis kernel based on genome-wide gene expression; and the pathway version, \texttt{GPR:M}, which uses kernels computed using genes from individual pathways and a genome-wide gene expression kernel computed using all genes not included in any of the pathways. 

We selected six pathways based on the existing knowledge of the molecular characteristics of kidney renal clear cell carcinoma (KIRC). The PI3K/AKT/mTOR pathway (\textit{PI3K}) was selected because genes in this pathway were recurrently mutated in KIRC patients \citep{cancer2013comprehensive}.  Because KIRC is characterized by a disordered metabolism, we also selected four pathways associated with metabolic function: glycolysis and gluconeogenesis (\textit{Glyc/Gluc}), metabolism of fatty acids (\textit{Fatty acid}), pentose phosphate pathway (\textit{Pent/Phos}), and citrate cycle pathway (\textit{Citrate cycle}). The genes belonging to three pathways \textit{PI3K}, \textit{Glyc/Gluc}, and \textit{Fatty acid} were obtained from Molecular Signatures Database hosted by the Broad Institute \href{http://software.broadinstitute.org/gsea/msigdb}{http://software.broadinstitute.org/gsea/msigdb}, and the genes of pathways \textit{Pent/Phos} and \textit{Citrate cycle} were obtained from Pathway Commons \href{http://www.pathwaycommons.org/}{http://www.pathwaycommons.org/}. Finally, we included the pathway/gene-set, which includes
genes that are expressed in different immune cell types and are used in the CIBERSORT software package \citep{newman2015robust} (\textit{CSORT}). We included this gene set to capture potential immune-related information because previous studies have associated long survival time in KIRC patients with immune infiltration \citep{escudier2012emerging}.

We use our proposed confounder adjustment approach to adjust for the potential confounding of sequencing plate ID, age, and tumor stage. Specifically, we fit the multivariate regression of gene expression on sequencing plate ID, tumor stage, age, and gender, and compute the candidate kernels using the residuals. We found this adjustment had a minimal effect on survival time prediction accuracy relative to the version without confounding adjustment.

\subsection{Analyses and results}
Following the setup of our simulation studies, we randomly split the data into training and testing sets of size 413 and 100, respectively, for 500 independent replications. To compare the performance across methods, we use the same metrics defined in Section \ref{sec:metrics}. Unlike our simulation studies, the test set contains censored survival times, so when computing the C-index and integrated Brier score, we estimate $\hat{H}$ using the Kaplan-Meier estimator on the testing data.  The best C-index average was 0.746, the best integrated AUC average was 0.794, and the best integrated Brier score average was 0.155. 

We display results for a subset of the competitors considered in our simulation studies in Figure \ref{fig:Analysis_results}. Both variations of our proposed method, \texttt{GPR:K} and \texttt{GPR:M}, significantly outperformed the competitors across all three metrics. We noticed that the genome-wide version of our estimator tended to outperform the estimator which used pathway-based kernels. To further illustrate the contribution of each variance component in pathway-based analysis, we display boxplots of the relative magnitudes of the estimated variance components across the five hundred replications (Figure \ref{fig:Analysis_results}(d)). We found that genome-wide effects accounted for approximately 70\% of variability in log-survival time, whereas random noise accounted for 25\%-30\%. Of the considered pathways, the pentose phosphate and fatty acid metabolism pathways were most frequently estimated to be away from zero, with the fatty acid metabolism pathway contributing nearly 50\% of variability in a small number of replications. Overall these results suggest that these few pathways do not make a significant contribution to survival time prediction and most information are from genome-wide gene expression.

\section{Discussion}
In this paper, we propose a new method to predict survival time using genome-wide gene expression data. Using the framework of Gaussian process regression, we develop a flexible and computationally efficient algorithm to perform two tasks: to impute censored survival time and to estimate the parameters of the model. We model the covariance structure of the log-survival time using one or more kernels defined using gene expression data. In both simulations and real data analyses, we define multiple kernels using gene expression from multiple pathways, though in practice, these kernels can be defined using the same set of genes with different definitions of distance/similarities, or they can be defined based on multiple types of -omic data. Although we have developed our method for survival time prediction, it can be used or extended for other outcomes with certain patterns of missing data. 

There are several directions to improve or extend our method. When the number of kernels is large, e.g., tens or hundreds of kernels, some regularization or penalty should be applied on the weight of each kernel. Although our simulations studies have demonstrated that our method is not sensitive to deviation from Gaussian distribution assumption, it is desirable to develop more robust non-parametric or semi-parametric approaches \citep{zeng2007maximum}. 

\bibliography{arXiv}

\begin{thebibliography}{}

\bibitem[Banerjee et~al., 2003]{banerjee2003frailty}
Banerjee, S., Wall, M.~M., and Carlin, B.~P. (2003).
\newblock Frailty modeling for spatially correlated survival data, with
  application to infant mortality in {M}innesota.
\newblock {\em Biostatistics}, 4(1):123--142.

\bibitem[Bender et~al., 2005]{bender2005generating}
Bender, R., Augustin, T., and Blettner, M. (2005).
\newblock Generating survival times to simulate {C}ox proportional hazards
  models.
\newblock {\em Statistics in Medicine}, 24(11):1713--1723.

\bibitem[Caffo et~al., 2005]{caffo2005ascent}
Caffo, B.~S., Jank, W., and Jones, G.~L. (2005).
\newblock Ascent-based {M}onte {C}arlo expectation--maximization.
\newblock {\em Journal of the Royal Statistical Society: Series B (Statistical
  Methodology)}, 67(2):235--251.

\bibitem[{Cancer Genome Atlas Research Network}, 2013]{cancer2013comprehensive}
{Cancer Genome Atlas Research Network} (2013).
\newblock Comprehensive molecular characterization of clear cell renal cell
  carcinoma.
\newblock {\em Nature}, 499(7456):43.

\bibitem[Datta, 2005]{datta2005estimating}
Datta, S. (2005).
\newblock Estimating the mean life time using right censored data.
\newblock {\em Statistical Methodology}, 2(1):65--69.

\bibitem[Datta et~al., 2007]{datta2007predicting}
Datta, S., Le-Rademacher, J., and Datta, S. (2007).
\newblock Predicting patient survival from microarray data by accelerated
  failure time modeling using partial least squares and {LASSO}.
\newblock {\em Biometrics}, 63(1):259--271.

\bibitem[Deng et~al., 2016]{deng2016predicting}
Deng, D., Du, Y., Ji, Z., Rao, K., Wu, Z., Zhu, Y., and Coley, R.~Y. (2016).
\newblock Predicting survival time for metastatic castration resistant prostate
  cancer: {An} iterative imputation approach.
\newblock {\em F1000Research}, 5.

\bibitem[Escudier, 2012]{escudier2012emerging}
Escudier, B. (2012).
\newblock Emerging immunotherapies for renal cell carcinoma.
\newblock {\em Annals of Oncology}, 23.

\bibitem[Fern{\'a}ndez et~al., 2016]{fernandez2016gaussian}
Fern{\'a}ndez, T., Rivera, N., and Teh, Y.~W. (2016).
\newblock Gaussian processes for survival analysis.
\newblock In {\em Advances in Neural Information Processing Systems}, pages
  5021--5029.

\bibitem[Flegal et~al., 2017]{mcmcse}
Flegal, J.~M., Hughes, J., Vats, D., and Dai, N. (2017).
\newblock {\em mcmcse: {M}onte {C}arlo standard errors for {MCMC}}.
\newblock Riverside, CA, Denver, CO, Coventry, UK, and Minneapolis, MN.
\newblock R package version 1.3-2.

\bibitem[G{\"o}nen and Alpayd{\i}n, 2011]{gonen2011multiple}
G{\"o}nen, M. and Alpayd{\i}n, E. (2011).
\newblock Multiple kernel learning algorithms.
\newblock {\em Journal of Machine Learning Research}, 12(Jul):2211--2268.

\bibitem[Horrace, 2005]{horrace2005some}
Horrace, W.~C. (2005).
\newblock Some results on the multivariate truncated normal distribution.
\newblock {\em Journal of Multivariate Analysis}, 94(1):209--221.

\bibitem[Klein et~al., 1999]{klein1999modeling}
Klein, J.~P., Pelz, C., and Zhang, M.-J. (1999).
\newblock Modeling random effects for censored data by a multivariate normal
  regression model.
\newblock {\em Biometrics}, 55(2):497--506.

\bibitem[Newman et~al., 2015]{newman2015robust}
Newman, A.~M., Liu, C.~L., Green, M.~R., Gentles, A.~J., Feng, W., Xu, Y.,
  Hoang, C.~D., Diehn, M., and Alizadeh, A.~A. (2015).
\newblock Robust enumeration of cell subsets from tissue expression profiles.
\newblock {\em Nature Methods}, 12(5):453.

\bibitem[Owen, 2017]{owen2017statistically}
Owen, A.~B. (2017).
\newblock Statistically efficient thinning of a {M}arkov chain sampler.
\newblock {\em Journal of Computational and Graphical Statistics},
  26(3):738--744.

\bibitem[Uno et~al., 2011]{uno2011c}
Uno, H., Cai, T., Pencina, M.~J., D'Agostino, R.~B., and Wei, L. (2011).
\newblock On the {C}-statistics for evaluating overall adequacy of risk
  prediction procedures with censored survival data.
\newblock {\em Statistics in Medicine}, 30(10):1105--1117.

\bibitem[Uno et~al., 2007]{uno2007evaluating}
Uno, H., Cai, T., Tian, L., and Wei, L. (2007).
\newblock Evaluating prediction rules for t-year survivors with censored
  regression models.
\newblock {\em Journal of the American Statistical Association},
  102(478):527--537.

\bibitem[Van~Wieringen et~al., 2009]{van2009survival}
Van~Wieringen, W.~N., Kun, D., Hampel, R., and Boulesteix, A.-L. (2009).
\newblock Survival prediction using gene expression data: a review and
  comparison.
\newblock {\em Computational Statistics \& Data Analysis}, 53(5):1590--1603.

\bibitem[Wei and Tanner, 1990]{wei1990monte}
Wei, G.~C. and Tanner, M.~A. (1990).
\newblock A {M}onte {C}arlo implementation of the {EM} algorithm and the poor
  man's data augmentation algorithms.
\newblock {\em Journal of the American Statistical Association},
  85(411):699--704.

\bibitem[Wheeler et~al., 2014]{wheeler2014poly}
Wheeler, H.~E., Aquino-Michaels, K., Gamazon, E.~R., Trubetskoy, V.~V., Dolan,
  M.~E., Huang, R.~S., Cox, N.~J., and Im, H.~K. (2014).
\newblock Poly-omic prediction of complex traits: {OmicKriging}.
\newblock {\em Genetic Epidemiology}, 38(5):402--415.

\bibitem[Wilhelm and Manjunath, 2015]{tmvtnorm}
Wilhelm, S. and Manjunath, B. (2015).
\newblock {\em {tmvtnorm}: {T}runcated multivariate normal and {S}tudent t
  distribution}.
\newblock R package version 1.4-10.

\bibitem[Wu et~al., 1983]{wu1983convergence}
Wu, C.~J. et~al. (1983).
\newblock On the convergence properties of the {EM} algorithm.
\newblock {\em The Annals of Statistics}, 11(1):95--103.

\bibitem[Wu et~al., 2008]{wu2008method}
Wu, T., Sun, W., Yuan, S., Chen, C.-H., and Li, K.-C. (2008).
\newblock A method for analyzing censored survival phenotype with gene
  expression data.
\newblock {\em BMC Bioinformatics}, 9(1):417.

\bibitem[Zeng and Lin, 2007]{zeng2007maximum}
Zeng, D. and Lin, D. (2007).
\newblock Maximum likelihood estimation in semiparametric regression models
  with censored data.
\newblock {\em Journal of the Royal Statistical Society: Series B (Statistical
  Methodology)}, 69(4):507--564.

\bibitem[Zhou et~al., 2015]{zhou2015mm}
Zhou, H., Hu, L., Zhou, J., and Lange, K. (2015).
\newblock {MM} algorithms for variance components models.
\newblock {\em arXiv preprint arXiv:1509.07426}.

\bibitem[Zhu et~al., 2017]{zhu2017integrating}
Zhu, B., Song, N., Shen, R., Arora, A., Machiela, M.~J., Song, L., Landi,
  M.~T., Ghosh, D., Chatterjee, N., Baladandayuthapani, V., et~al. (2017).
\newblock Integrating clinical and multiple omics data for prognostic
  assessment across human cancers.
\newblock {\em Scientific Reports}, 7(1):16954.

\bibitem[Zhu and Kosorok, 2012]{zhu2012recursively}
Zhu, R. and Kosorok, M.~R. (2012).
\newblock Recursively imputed survival trees.
\newblock {\em Journal of the American Statistical Association},
  107(497):331--340.

\end{thebibliography}

\end{document}